\documentclass{elsart} 
\usepackage{graphicx,amssymb}
\journal{Nuclear Physics A}
\newcommand{\tr}{{\rm tr}}

\begin{document}
\begin{frontmatter}
\title{Instantons As Unitary Spin Maker}

\author[Leon]{M.\ Napsuciale},
\ead{mauro@feynman.ugto.mx}
\author[Juelich]{A.\ Wirzba\corauthref{cor}},
\corauth[cor]{Corresponding author.}
\ead{a.wirzba@fz-juelich.de}
\author[Zacatecas]{M.\ Kirchbach}
\ead{\\kirchbach@chiral.reduaz.mx}
\address[Leon]{Instituto de Fisica, Universidad de Guanajuato,\\
AP E-143, 37150, Leon, Guanajuato, Mexico}
\address[Juelich]{Institut f\"ur Kernphysik (Theorie), FZ J\"ulich,\\
D-52425 J\"ulich, Germany}
\address[Zacatecas]{Escuela de Fisica, Univ. Aut. de Zacatecas,\\
AP C-580, Zacatecas, ZAC 98068 Mexico}

\begin{abstract}
We investigate the relevance 
of the instanton--induced determinantal 't Hooft interaction 
to the $\eta $--nucleon coupling $g_{\eta NN}$
within the framework of a three-flavor linear sigma model
in the OZI--rule--respecting basis
$\lbrace (\bar ss),{1\over \sqrt{2}}(\bar u u +\bar d d)\rbrace $.
Instantons, in combination with the spontaneous breaking of chiral symmetry,
are shown to provide the major mechanism for the ideal mixing between 
pseudoscalar strange and non-strange quarkonia.
As long as 't Hooft's interaction captures most of the
basic features of the axial QCD gluon anomaly, 
we identify the
anomaly as the main culprit
for the appearance of octet flavor symmetry in the anomalous
sectors of the pseudoscalar (and axial vector) mesons.
Within this context, unitary spin
is shown to be an accidental symmetry due to anomalous gluon dynamics 
rather than a fundamental symmetry in its own right. 
Though  we find the $\eta$--nucleon coupling constant
$g_{\eta NN}$ to obey a Goldberger-Treiman
like relation, the latter does not take its
origin from a pole dominance of the induced pseudoscalar
form factor of the octet axial current, but from a subtle
flavor-mixing mechanism that is traced back to instanton dynamics.
The model presented allows for possible generalizations to non-ideal mixing 
angles and different values of the meson decay constants in the strange and
non-strange sectors, respectively. Finally, we discuss the issue as
to what extent the $\eta $ meson may be considered as
a Goldstone boson under the constraints of the anomaly-produced
unitary spin.
\end{abstract}
\begin{keyword}
flavor symmetry \sep eta meson \sep instantons \sep OZI rule \sep 
$\eta N$ coupling
\PACS  11.30.Hv \sep 11.40.Ha\sep 14.40.Aq
\end{keyword}
\end{frontmatter}

\section{Introduction}

Flavor symmetry is presently understood on the basis of 
QCD and the structure of the quark mass matrix. In the zero quark
mass limit, the light flavor sector of
QCD acquires a three flavor chiral symmetry 
$U(3)_L\times U(3)_R$ which at the level of hadrons is assumed to be 
realized in the Goldstone phase.
The associated Goldstone bosons
are identified with the lightest pseudoscalar mesons. 
 
The first problem one encounters in that scheme is the large mass of the 
$\eta^\prime $. The way out of this problem is to take into 
account quantum corrections which spoil the conservation of the singlet axial 
current. Particularly relevant to this problem is the existence of 
Euclidean solutions with non-trivial topological 
properties (instantons) which also break the U(1)$_A$ symmetry. In fact, 
the most appealing explanation of the problem of the large $\eta^\prime$ 
mass is provided by the instanton induced quark-quark interaction 
\cite{thooft1}. 

The second problem concerns the different
symmetries for pseudoscalar and axial vector mesons, on the one side,
and vector and tensor mesons, on the other side.

Isoscalar vector mesons closely follow the flavor basis 
structure: $\phi = \bar\psi\lambda_{\rm{s}}\psi = \bar s s$ and 
$\omega = \bar\psi\lambda_{\rm{ns}}\psi= (\bar u u + \bar dd)/\sqrt{2}$.
This contrasts the singlet-octet pattern followed by the isoscalar 
axial vector and pseudoscalar mesons.
To explain this dilemma, the Okubo-Zweig-Iizuka (OZI) rule 
\cite{Okubo,Zweig,Iizuka} was invoked that forbids the mixing of
of quarkonia made of $u$ or $d$ quarks with quarkonia made of $s$ or $c$
quarks.
In the absence of symmetry breaking terms one can freely use any basis 
for the generators of the group. The physically 
interesting basis is the one whose generators still reflect  
a residual symmetry of the system
 in the presence of symmetry breaking terms. In that regard,  
we have three sources of breaking  $U(3)_L\times U(3)_R$ symmetry. 
The first one is the 
$U_A(1)$ symmetry breaking by quantum effects. Although the
instanton induced interaction is suppressed at high energies 
due to the factor $\exp(-8\pi^2/g^2)$, it is always present and 
even becomes decisive at low energies. The second source is the 
spontaneous 
breaking of chiral symmetry occurring at the scale 
$\lambda_\chi \approx 4\pi f_{\pi}$. 
The third source is the non-vanishing  quark mass matrix that explicitly
violates three flavor chiral symmetry.

Isospin is a good symmetry for hadronic strong interactions, and in the 
isospin limit  the quark mass matrix  
simplifies to  $M_q$ = Diag($m$,$m$,$m_s$). This matrix can be written 
in terms of the $U(3)$ (not of $SU(3)$) generators. Using $SU(3)$ 
would require to consider simultaneously  two irreducible representations, 
the singlet and the octet.

The problem of the conflicting flavor symmetries
of the anomalous- and anomaly-free meson sectors was addressed
e.g.\ in Ref.~\cite{MKPRD}. There, the (three) flavor symmetry 
for hadrons was concluded from the conserved vector current 
rather than from mass relations and shown to be
 $SU(2)_{ud}\otimes SU(2)_{cs}\otimes U(1_{udcs})$  in 
the limit of 
heavy {\em  spectator} $c$ quarks.  Apparently, this symmetry respects, from 
the very beginning, the quark generations and the OZI rule. 
Within the framework of $SU(2)_{ud}\otimes SU(2)_{cs}\otimes U(1_{udcs})$,
the $SU(3)$ flavor symmetry for the pseudoscalar and axial vector mesons
appears as an artifact of the axial $U(1)_A$ gluon anomaly.

It is the goal of the present study to show that unitary spin is an 
accidental symmetry (in the language of Ref.~\cite{Holger}) 
that is manufactured by the instanton dynamics, 
{\em and}  to explore the implications for the $\eta N$ coupling.

In the next section we shall illustrate the formation of the
``eightfold way'' within a linear sigma model which has 't Hooft' s
determinantal flavor--dependent interaction built in.  We will analyze
the resulting consequences for the $\eta NN$ coupling 
constant in Section~\ref{coupling}. The paper ends with a brief summary.

\section{``Eightfold Way'' from  instanton dynamics.}
\label{eightfold}

Let us first state the notation for
the wave function of the $\eta$:
\begin{eqnarray}
 |\eta\rangle &=&\cos \theta_P\,|\eta^8\rangle 
 -\sin\theta_P \,|\eta ^1\rangle \; ,\label{etaoct}\\
 |\eta ^8\rangle &=& {1\over \sqrt{6}}
 (\bar u u + \bar d d  -2\bar ss)\;, \qquad 
 |\eta ^1 \rangle =  {1\over \sqrt{3}}(\bar u u +\bar dd +\bar ss) \; .
 \label{eta_oct}
\end{eqnarray}
Here, $|\eta^8\rangle $ and $|\eta^1\rangle $ are in turn the
Gell-Mann's octet and singlet $ 0^{-+}$
states, respectively. 
The mixing angle $\theta_P$ is obtained
on the grounds of quadratic ($\theta_P\approx -10.1^\circ $), or 
linear ($\theta_P \approx -20^\circ  $) meson-mass formulae~\cite{DonGolHol92}.
Obviously, the non-strange quarkonium state $|\eta ^{\rm ns}\rangle $ 
is recovered  in Eq.\,(\ref{etaoct}), when the mixing 
angle  takes the ``ideal'' value of $\theta_P=\theta_{\rm{id}}$ \cite{PDG}
with
 $\theta_{\rm{id}}\approx -54.7^\circ $  
($\cos\theta_{\rm{id}}=1/\sqrt{3} ~,~ \sin\theta_{\rm{id}}=-\sqrt{2/3}$ ). 

Alternatively, in the OZI (generation) basis, the $\eta $ wave 
function takes the form~\footnote{We adopt the following convention:
mixing angles of a meson $M$ in the octet--singlet basis are
denoted as $\theta_M$ and in the non-strange--strange basis as
$\phi_M$.}
\begin{eqnarray}
 &&|\eta\rangle =\cos \phi_P\,|\eta^{\rm {ns}}\rangle 
 -\sin\phi_P \,|\eta ^{\rm {s}}\rangle \; ,
 \label{etawafu}\\
 &&|\eta ^{\rm {ns}}\rangle = {1\over \sqrt{2}}
 \left(\bar u u + \bar d d\right)\;, \qquad 
 |\eta ^{\rm {s}} \rangle =  \bar s s \; .
 \label{eta_wafu}
\end{eqnarray}
A value of $\phi_P \approx 39.4^\circ $ has recently been 
concluded from fitting $\eta (\eta ')$ decay cross 
sections \cite{Feldmann1,Feldmann2}.
The $\eta^8$ state can be recovered
from Eq.\,(\ref{etawafu}) if $\phi_P=\phi_{\rm{su(3)}}
\approx 54.7^\circ$ ($\sin \phi_{\rm{su(3)}}= \sqrt{2/3}~,~ 
\cos\phi_{\rm{su(3)}}=\sqrt{1/3}$) is inserted instead. 
The angles in the two bases are 
related as $\theta_P = \phi_P - 54.7^\circ $ in 
such a way that the physical mixing angle $\phi_P \approx 39.4^\circ $ in
the OZI basis corresponds to the value
$\theta _P \approx -15.3^\circ $  in the octet-singlet basis and
thereby lies between the quadratic
and linear mixing angles.

To study the formation of the ``eightfold way''
in an explicit framework, we introduce here 
a chirally symmetric $ \lbrack U(3)_L\otimes U(3)_R\rbrack $
lagrangian  for a scalar and a pseudoscalar mesonic
nonet, in turn denoted by $(\sigma_i)$ and $(P_i)$,
\begin{equation}
{\mathcal L}= {\mathcal L}_{sym}  + {\mathcal L}_{U_A(1)} + 
{\mathcal L}_{SB}\;, \label{L-model}
\end{equation}
where the lagrangian $ {\mathcal L}_{sym}$ describes the 
flavor-symmetric part, the lagrangian  $ {\mathcal L}_{U_A(1)}$ the 
flavor-breaking from the $U_A(1)$ gluon anomaly, and
the lagrangian ${\mathcal L}_{SB}$ the explicit symmetry breaking.

The  flavor-symmetric lagrangian 
$ {\mathcal L}_{sym}$ is given by 
\begin{eqnarray}\nonumber
 {\mathcal L}_{sym}&=&\frac{1}{2}\tr\left[(\partial_\mu M)
 (\partial^\mu M^{\dagger}) \right]  -\frac{\mu^2}{2} 
 X (\sigma ,P)\\
 && \mbox{} -\frac{\lambda}{4} Y(\sigma ,P)
 -\frac{\lambda^\prime}{4} X^2 (\sigma , P) \; , 
 \label{symmlag}
\end{eqnarray}
where
$M\equiv\sigma + i P$, and $X,Y$ stand in turn for the left-right
symmetric traces 
\begin{equation}\begin{array}{lclcl}
 X(\sigma,P)&\equiv&\tr\left[ M M^{\dagger}\right]  
  &=& \tr\left[\sigma^2+P^2\right]\;,\\
 Y(\sigma,P)&\equiv&\tr\left[ (M M^{\dagger})^2\right ]
 &=& \tr\left[ (\sigma^2+P^2)^2+2\left(\sigma^2 P^2-(\sigma P)^2\right)
 \right]\; .
\end{array}  
\end{equation}The pseudoscalar and
scalar matrix fields $P$ and $\sigma $ are written in terms of a specific
basis spanned by seven of the standard Gell-Mann matrices, namely $\lambda_i
~(i=1,\ldots , 7)$, {\em and} by two unconventional matrices 
$\lambda_{\rm{ns}}$=diag(1,1,0), and $\lambda_{\rm{s}}$ = $\sqrt{2}$
diag(0,0,1), respectively.  The decomposition obtained in this way reads
$P\equiv \frac{1}{\sqrt{2}} \lambda_i P_i$ with 
$i=ns,s,1,\ldots,7$ and similarly
for the scalar field. 

As we will show below, 
the main effect in furnishing the masses of the isoscalar-pseudoscalar 
mesons comes from the axial $U_A(1)$ gluon anomaly. 
There exist two  candidates 
for the phenomenological description of the
above  effect that are associated with two 
different contributions to ${\mathcal L}$ 
that lead to $U(1)_A$ symmetry breaking at the level of QCD. 
The first one is the flavor determinantal interaction 
\begin{equation}
 {\mathcal L}_{U_A(1)} = {\mathcal L}_{INST}=- \beta Z(\sigma,P) \; ,
\label{lins} 
\end{equation}
where
\begin{eqnarray} 
 Z(\sigma ,P) &=& \lbrace  \mbox{det}(M)+ \mbox{det}
 (M^{\dagger }) \rbrace \nonumber \\
  &=& \frac{1}{3}\left\{ \tr[\sigma]
 \left( \left(\tr[ \sigma] \right)^2-3\left(\tr [P]\right)^2\right)
 -3\tr[\sigma]\,\tr[\sigma^2-P^2] \right.\nonumber  \\
 && \quad\left.\mbox{}
 + 6 \tr[ P]\, \tr[\sigma P]
 +2 \tr\left[
  \sigma\left(\sigma^2-3 P^2\right)\right]\right\}\;.
 \label{instanton} 
\end{eqnarray}
It stands for the bosonized  't Hooft effective quark-quark 
interaction which is induced by instantons (gluon configurations with 
integer topological charge) and 
turns out to be  determinantal and flavor dependent~\cite{thooft1,thooft2}.
The other  candidate is the Veneziano-Witten interaction term
\begin{equation}
{\mathcal L}_{U_A(1)} = {\mathcal L}_{VW} = \beta^\prime\, VW(\sigma, P)
\end{equation}
\noindent where 
\begin{equation}
VW(\sigma,P)=
\left( \ln \det (M)- \ln \det 
 (M^{\dagger }) \right) ^2 ,
\end{equation}
\noindent which takes its origin from the fluctuations of the 
topological charge 
(the ``ghost pole'' mechanism) \cite{veneziano,witten,schechter,arnowitt}. 
Consequences of this mechanism were investigated at the level of the 
fundamental theory in \cite{diakonov} where axial Ward identities were 
exploited to extract information on pseudoscalar mesons. 
However,
there exists a relation between these two mechanisms for the 
$U_A(1)$-breaking. 
Indeed, by
assuming a non-vanishing 
vacuum expectation value of $M$ and
expanding in powers of  
the ``fluctuation'' 
$\Delta(\det M)=\det M-\langle \det M\rangle $ 
the authors of Ref.\,\cite{schechter} showed 
that
both effective interactions can be linked as
\begin{equation}
VW
-\Bigl({\det\langle M\rangle}\Bigr)^{-2}\,Z^2= U(3)_L\otimes U(3)_R 
~{\rm invariant~ operator}.\label{ins-vw}
\end{equation}
The first term on the l.h.s. of Eq.\,(\ref{ins-vw}) can even be 
related to $Z$ itself  by a further expansion  in powers of
the fluctuation 
\cite{schechter}. 
Clearly, these are not the only possibilities
 for the breaking $U_A(1)$ symmetry. Other possibilities arising from 
different approximations in the calculational schemes 
of instanton effects have been proposed \cite{diakonov2}. 
In the present work we restrict ourselves to
the usage of the  effective 't Hooft interaction. 
At the phenomenological level it is
a promising candidate not only  for providing a
mechanism of the flavor-mixing but also for 
simultaneously explaining the unusual properties of scalar 
mesons, including 
the long standing problem of the two-photon decays of the $a_0(980)$ and 
$f_0(980)$ \cite{luna,nalu}.

Finally, the model lagrangian (\ref{L-model}) contains  the standard term
\begin{equation}
 {\mathcal L}_{SB} =  \tr \left[ c\sigma \right] \label{lbreak}
\end{equation}
which breaks the left-right symmetry explicitly.
The $c$ matrix is spanned by the same basis 
$c\equiv \frac{1}{\sqrt{2}} \lambda_i c_i$ as the fields,
and the nine expansion coefficients $c_i$ are independent 
constants. The most general  $c$-matrix that  preserves isospin,
respects PCAC and is consistent with the  quark mass matrix, 
has  $c_{\rm{s}}$ and $c_{\rm{ns}}$ as the only non-vanishing 
entries. While the $c_{\rm{s}}$ term  explicitly causes  
$\lbrack U(3)_L\times U(3)_R\rbrack /\lbrack U(2)_L\times U(2)_R\rbrack$
breaking, the $c_{\rm{ns}}$ leads to 
$\lbrack U(2)_L\otimes U(2)_R\rbrack/U(2)_I$ breaking.  
{}Furthermore, the linear $\sigma$  term in Eq.\,(\ref{lbreak}) induces 
$\sigma$-vacuum transitions which supply the scalar
fields  with non-zero vacuum expectation values (v.e.v) 
(hereafter denoted by $\langle \cdots\rangle $). 
To simplify notations, let us re-denote 
$\langle \sigma \rangle $ by $V$ with $V= $diag ($a,a,b$), where  
$a$ and $b$ denote the vacuum expectation values
of the strange and non-strange quarkonium, respectively,
\begin{equation}
 a=\frac{1}{\sqrt{2}}\langle \sigma_{\rm{ns}}\rangle\;, \qquad
 b=\langle \sigma_{\rm{s}}\rangle \;. \label{vev}
\end{equation}
We now shift, as usual, the old $\sigma $ field to a  new scalar
field
$S=\sigma -V$ such that $\langle S\rangle=0$. In this way, 
new  mass terms, three-meson interactions,
and a linear term are generated. In particular, the mass and linear terms
read
\begin{eqnarray} 
 {\mathcal L}_2 &=& -\frac{1}{2} \left( \mu^2  +\lambda'( 2 a^2 + b^2) 
 -2 \lambda a b \right)\tr[S^2 +P^2] \nonumber\\
 && \mbox{}- 
 \lambda (a+b)\,\tr\left[ V\left(S^2+P^2\right)\right] 
  - \frac{\lambda }{2} \,\tr\left[ (VS)^2 -(VP)^2 \right] \nonumber \\
&&\mbox{} +\beta(2a+b)\, \tr[S^2-P^2]
 -2\beta \,\tr\left[V\left(S^2-P^2\right)\right] \nonumber \\
 && \mbox{} 
 -\beta (2 a+b) 
 \left\{ \left(\tr[S]\right)^2\mbox{$-$}\left(\tr[P]\right)^2 \right\}
 +2\beta \left( \tr[S]\, \tr[VS]\mbox{$-$}\tr[P]\,\tr[VP]\right) 
 \nonumber \\
 && \mbox{}
 - \lambda' \left( \tr\left[VS \right]\right)^2 \,,
  \label{Lmass2} \\
 {\mathcal L}_1 &=& \tr[c S] 
 -\left\{\mu^2 + \lambda'(2a^2+b^2) + \lambda(a^2+ab+b^2)
 -2\beta a \right\} 
 \,\tr[SV] \nonumber \\
&&\mbox{}
 +  a (a+b)\left(\lambda b   - 2 \beta \right) \,\tr[S] \,.
 \label{Llin}
\end{eqnarray}
The above mentioned  terms are affected -- via the
't Hooft determinant (see the terms proportional to the parameter $\beta$) -- 
by  the $U_A(1)$ anomaly which couples 
to the v.e.v' s of the scalar fields by the spontaneous breaking 
of chiral symmetry. The consequence 
is the breaking of the original symmetry 
down to $SU(2)_I$ isospin. 
A detailed analysis of the  breaking pattern 
for different values of the parameters
and the corresponding implications for Goldstone modes 
in this model was carried in \cite{lenaghan}. We refer the interested reader 
to this work for further details.

The masses of the seven unmixed pseudoscalar and scalar 
mesons corresponding to the original Gell-Mann matrices
$\lambda_i$ ($i=1,\ldots,7$), namely 
the {\em isovector} pseudoscalar ($\pi$) and 
scalar ($a_0 $) mesons as well as 
the two {\em isodoublets} of 
pseudoscalar  ($K$) and scalar ($\kappa $) 
mesons,
are obtained from the first five terms  of the second-order lagrangian 
(\ref{Lmass2})
as 
\cite{thooft3,napsu,tornq} 
\begin{eqnarray} 
 \begin{array}{lclclcl}
 m^2_\pi &=&\xi+2\beta b+\lambda a^2 \; , &\ \ &
 m^2_{a_0} &=& \xi-2\beta b +3\lambda a^2    \; ,\\
 m^2_K &=&\xi+2\beta a+\lambda (a^2\mbox{$-$}ab\mbox{+}b^2)\;,  &\ \ &
 m^2_\kappa &=&\xi-2\beta a+\lambda (a^2\mbox{$+$}ab\mbox{+}b^2)
\; , \end{array}
 \label{unmixedmasses}
\end{eqnarray}
where we used the convenient short--hand notation\footnote{In 
the pseudoscalar sector, the ``bare mass'' $\mu$  solely appears
inside  this
combination.} 
$\xi \equiv \mu^2 +\lambda^\prime(2a^2+b^2)$. 
The elimination of the linear terms of the first-order lagrangian (\ref{Llin})
imposes the following constraints on the explicit-symmetry-breaking terms
$c_{\rm{ns}}$, and $c_{\rm{s}}$:
\begin{equation}
 c_{\rm{ns}} = \sqrt{2} a m^2_\pi\;,\qquad
 c_{\rm{s}} = b m_K^2 + a(m_K^2 - m_\pi^2)\; .
 \label{c-equ-set}
\end{equation}
In Ref.\cite{napsu} the  PCAC relations for the pion and kaon field are
discussed. Whereas the PCAC relation of the pion can be directly read off
the $c_{\rm ns}$ term, the one of the kaon has to be inferred from
the 
linear combination 
 $ \left(c_{\rm s} + c_{\rm ns}/\sqrt{2}\right)/\sqrt{2}$:
\begin{equation}
 f_\pi =\sqrt{2} a \;, \qquad
 f_K = \frac{1}{\sqrt{2}} (a+b)\;.\label{fkfpi}
\end{equation}
The mass term of the lagrangian involving the mixed {\em   isoscalar}
pseudoscalar and scalar  fields, which correspond
to the  unconventional $\lambda_{\rm ns}$ and $\lambda_{\rm s}$
matrices, gets in addition to the contributions from the first five terms also 
contributions from the last three terms of (\ref{Lmass2})
and reads  therefore
\begin{eqnarray} 
 {\mathcal L}_{mass} &=&-\frac{1}{2} (m^2_{P_{\rm ns}} P^2_{\rm ns}+ 
 m^2_{P_{\rm s}} P^2_{\rm s} + 
 2m^2_{P_{\rm s-ns}} P_{\rm s} P_{\rm ns})\nonumber \\
 &&
 -\frac{1}{2} (m^2_{S_{\rm ns}} S^2_{\rm ns}+ 
 m^2_{S_{\rm s}} S^2_{\rm s} + 
 2m^2_{S_{\rm s-ns}} S_{\rm s} S_{\rm ns})
 \label{Lmass_mix}
\end{eqnarray}
with
\begin{eqnarray}
  m^2_{P_{\rm{ns}}} =\xi-2\beta b + \lambda a^2 \;,  &\qquad &
  m^2_{S_{\rm {ns}}} =\xi+2\beta b + 3\lambda a^2 +4\lambda^\prime a^2\;,
 \label{snsmixing1}
 \\
 m^2_{P_{\rm {s}}} =\xi + \lambda b^2 \; ,  & \qquad & 
 m^2_{S_{\rm{s}}} =\xi + 3\lambda b^2 + 2\lambda^\prime b^2 
 \label{snsmixing2}\; ,
 \\ 
 m^2_{P_{\rm {s-ns}}} = -2\sqrt{2}\beta a \; , & \qquad & 
 m^2_{S_{\rm{s-ns}}} = 2\sqrt{2}(\beta+\lambda^\prime b)a \;.
 \label{snsmixing3}
\end{eqnarray}
Here, $m_{\chi_{\rm s}}$ and $m_{\chi_{\rm{ns}}}$ with
$\chi\in\{P,S\}$
are  the masses of the 
strange and non-strange \mbox{(pseudo-)}scalar 
quarkonia respectively, while $m^2_{\chi_{\rm s-ns}}$, which
does not need to be positive, 
denotes the   transition mass-matrix elements
of the strange--non-strange (pseudo-)scalar 
quarkonia. 
Equations (\ref{snsmixing3}) show that the mixing
between strange and non-strange quarkonia is due to
the instanton-induced interactions {\em and}  
the spontaneous breakdown of chiral symmetry.

In the following we will first discuss the mixed {\em pseudoscalar} sector.
The physical isoscalar pseudoscalar 
fields are linear combinations of $ P_{\rm {s}}$, 
$P_{\rm {ns}}$ which 
diagonalize the pseudoscalar part of ${\mathcal L}_{mass}$: 
\begin{eqnarray} 
 \eta &=& P_{\rm{ns}}~ \cos \phi_P - P_{\rm{s}}~ \sin \phi_P \; ,\\
 \eta^\prime &=& P_{\rm{ns}}~ \sin \phi_P
 + P_{\rm{s}}~ \cos \phi_P \; . 
 \nonumber
\end{eqnarray}
This diagonalization of the mass matrix for the pseudoscalar mesons
yields the  relations 
\begin{eqnarray} 
 \sin ~2\phi_P = \frac{2 m^2_{P_{\rm{s-ns}}}}
 {m^2_{\eta^{\prime}}-m^2_{\eta}}\; ,  &\qquad &
 \cos ~2\phi_P = \frac{ m^2_{P_{\rm{s}} }-
 m^2_{P_{\rm{ns}} } }
 {m^2_{\eta^{\prime}}-m^2_{\eta}}\; .
 \label{mixing}
\end{eqnarray}
Here, $\phi_P$ 
stands for the isoscalar-pseudoscalar 
mixing angle as introduced in Eq.~(\ref{etawafu}) above.
In addition, one finds the following  trace relation 
\begin{equation}
 m^2_{\eta^\prime}+ m^2_{\eta} = m^2_{P_{\rm{s}}} + 
 m^2_{P_{\rm{ns}}} \label{trace}
\end{equation}
to be valid.
As a trivial consequence of Eq.\,(\ref{mixing}), the following relation holds:
$
(m^2_{\eta'}-m^2_\eta)^2= 4 m^4_{P_{\rm s-ns}}
+(m^2_{P_{\rm s}}-m^2_{P_{\rm ns}})^2 
$.  Together with Eq.\,(\ref{trace}) it induces
\begin{equation}
 m^2_{\eta'} m^2_\eta = m^2_{P_{\rm s}} m^2_{P_{\rm ns}} 
 - \left( m^2_{P_{\rm s-ns}}\right)^2
 \; , 
 \label{m2m2}
\end{equation}
such that
\begin{eqnarray} 
 m^2_{\eta/\eta'} 
 &=& \frac{1}{2}\left(m^2_{P_{\rm s}}+ m^2_{P_{\rm ns}}\right)
 \mp\sqrt{\frac{1}{4} \left( m^2_{P_{\rm s}}- m^2_{P_{\rm ns}} \right)^2
 +\left( m^2_{P_{\rm s-ns}}\right)^2 }\; . \label{meta2}
\end{eqnarray}
The five parameters entering the {\em pseudoscalar} sector of the
model ($\xi,~\lambda,~\beta,~a,~b$) can be fixed 
through the masses and the decay constants of the pseudoscalars 
($m_{\eta^\prime},~m_{\eta},~m_\pi,~m_K,~f_\pi$) and can be 
used to
predict all the other properties of the pseudoscalar mesons 
such as the mixing of the strange and non-strange fields
(see model\,1 of Table~\ref{input}). 
Alternatively, the kaon decay constant $f_K$ can be  used 
as input, replacing the 
combination ($m^2_{\eta^\prime}-m^2_{\eta}$) 
of the above given  quantities~\cite{tornq} (see model\,2 
of Table~\ref{input}
with the input 
($\sqrt{m_{\eta^\prime}^2+m_{\eta}^2},~m_\pi,~m_K,~f_\pi,~f_K $)).
The latter procedure creates slightly different results for the  
pseudoscalar mixing angle. Finally, one could also have  used 
the pseudoscalar mixing angle as input 
\cite{thooft3}
(see models 3a and 3b of Table~\ref{input}
with the input ($m_{\eta},~m_{\eta'},~m_\pi,~\theta_P,~f_\pi$)).
This leads to a different identification of 
the scalar nonet. The pertinent 
masses turn out to be highly sensitive to the 
choice for the input parameters. In particular, the latter 
version  yields heavy scalars~\cite{thooft3}. 

Now, by inserting (\ref{unmixedmasses}) for $m^2_{\pi}$, $m^2_K$
and (\ref{snsmixing1}--\ref{snsmixing3}) for $m^2_{P_{\rm ns}}$,
$m^2_{P_{\rm s}}$ , $m^2_{P_{\rm s-ns}}$ where the latter
are linked to $m^2_\eta$, $m^2_{\eta'}$ by the relations 
(\ref{trace}--\ref{m2m2}), one
can express
 the $\beta $ parameter 
in terms of the pseudoscalar meson masses according to
\begin{equation}
 -4\beta (a+b) = {(m^2_{\eta^\prime}-m^2_\pi )
 (m^2_\eta -m^2_\pi )\over (m^2_K -m^2_\pi )}\;. 
 \label{beta}
\end{equation}
The physical solution found here coincides with  the  phenomenologically
favored one of the two solutions of 
the  quadratic equation for $\beta $ reported in an earlier
work \cite{napsu}.
The parameter $a$ can be directly fixed through the first of 
Eqs.\,(\ref{fkfpi}),
whereas $b$, parameterized as $b=(1+2x)~a$, can be fixed either through  
a fit of the kaon mass as  $x=x_N=0.37$ \cite{napsu}, or directly
through $f_K$ in the second of Eqs.\,(\ref{fkfpi}) 
as $x=x_T=0.22$ \cite{tornq}.
Using, e.g., 
$x=0.37$ we obtain $\beta \approx - 1.55$\,GeV from (\ref{beta}) and
\begin{equation}
 \sin~2\phi_P = 0.9202\; , \qquad  \cos~2\phi_P =-0.3911  \label{sincos}
\end{equation}
from the mixing relations (\ref{mixing}) together with
Eq.\,(\ref{meta2}) 
and $m_{P_{\rm s}}^2< m_{P_{\rm ns}}^2$ (see Table~\ref{input}).
A careful analysis of the mass matrix shows that the actual mixing 
angle in the flavor basis is the one arising from the cosine 
relation in  Eq.\,(\ref{sincos}). This angle
is complementary to the one arising from the sine relation,
which does not distinguish between $\pi/2 -\phi_P $ and  $\phi_P $.
The mixing angle in the flavor basis thus turns out to lie within 
the range determined by the case $x=0.22$, namely  $\phi_P = 49.7^\circ $, 
on the one hand, and the case $x=0.37$, namely $\phi_P = 56.0^\circ $,
on the other hand (see Table~\ref{input}). The corresponding angle in the 
singlet-octet basis is $\theta_P = \phi_P-54.7^\circ$, where
$54.7^\circ $ results from the ideal mixing angle in the
 ns--s basis.
The so-determined values of $\theta_P$ are 
in the range $\theta_P~ \in [-5^\circ,~ +2^\circ]$ and  therefore close to 
zero
This finding, in combination with the fact that the sole 
mixing mechanism of
flavor fields is the instanton-induced interaction 
(see the left Eq.\,(\ref{snsmixing3})), establishes the main 
result of this section: 
{\it 't Hooft's instanton-induced 
interaction mixes strange and non-strange pseudoscalar fields 
in such a way that one of the physical fields becomes a member 
of the octet, while  the other one becomes 
an U(3) singlet}.  

The fact that models 3a and 3b  have a non-zero mixing angle
is in no contradiction to
this as the empirical value(s) of the mixing angle is used
as one of the
input parameters.  As discussed below, the small deviation from
the value zero (in comparison to the size of the ideal mixing) 
should be traced back to subleading contributions, as e.g.\ mesonic
loops.

If the coefficient $\lambda'$ in Eq.\,(\ref{snsmixing3}) is ignored, 
the mixing between the {\em scalar} strange and non-strange quarkonia
due to 't Hooft's instanton-induced interaction is predicted to be of the 
same size as the corresponding mixing in the pseudoscalar sector but with 
opposite sign. This is consistent with the results in 
\cite{schshur,isgur}.
In the {\em scalar sector\/}, however, one has
to account for the additional effect brought about by 
one of the 
chiral invariants in Eq.\,(\ref{symmlag}) whose strength is measured 
by the above-mentioned 
$\lambda^\prime$ coupling as dictated by the right relation of  
(\ref{snsmixing3}). 
As discussed in \cite{tornq}, this chiral invariant 
corresponds to OZI-rule violating disconnected hairpin diagrams.
They represent one out of various examples of subleading 
OZI-rule violating 
mechanisms, the most important among them being probably 
Lipkin's non-planar hadronic loops~\cite{Lip}:
in the scalar sector, the cancellation of 
hadronic loops  is strongly spoiled by 
parity conservation \cite{LipZou,Zou}. Thus, although subleading 
and suppressed with respect to instanton contributions, 
this effect acquires importance as it 
{\it interferes destructively with the instanton-induced 
contribution to the mixing of the scalar mesons}. 
This renders the scalars less strongly mixed than pseudoscalars and 
thus closer to the flavor basis. Estimates based on meson spectrum 
and on recent data on radiative $\phi$ decays involving scalars 
yield  for the isoscalar-scalar mixing angle 
$\phi_S \in [-9^\circ ,-14^\circ ]$ \cite{luna,nalu,napsu}. 
Therefore one of the scalar isoscalar mesons (sigma) can be nearly identified 
with the non-strange scalar and is strongly  moved down relative to the 
scalar isovector ($a_0$) as can be seen from 
Eqs.~(\ref{snsmixing1}, \ref{unmixedmasses}). This 
is also consistent with results in \cite{schshur}.

It is worth noting that the physical properties of mesons, belonging to
sectors which are not affected by the instanton-induced interactions, such as
the spin-1$^{--}$ and tensor $2^{++}$ mesons, are well described in terms of
almost pure flavor states.  The small departure from the flavor basis in these
sectors ($\phi_V \approx 4^\circ $) can be attributed to strong and yet
incomplete cancellations of {\em all} meson loops in this sector 
\cite{LipZou,Zou}.
The same argument can be used for the actual deviation of the $\eta $ from
being a pure octet state. The effect is slightly larger in that case due to
incomplete cancellations among hadronic loops \cite{Lip,LipZou,Zou}.

\section{Instanton dynamics and the  $\eta NN$ coupling}
\label{coupling}

\begin{figure}[htbp]
\begin{center}
\includegraphics*[width=7.5cm]{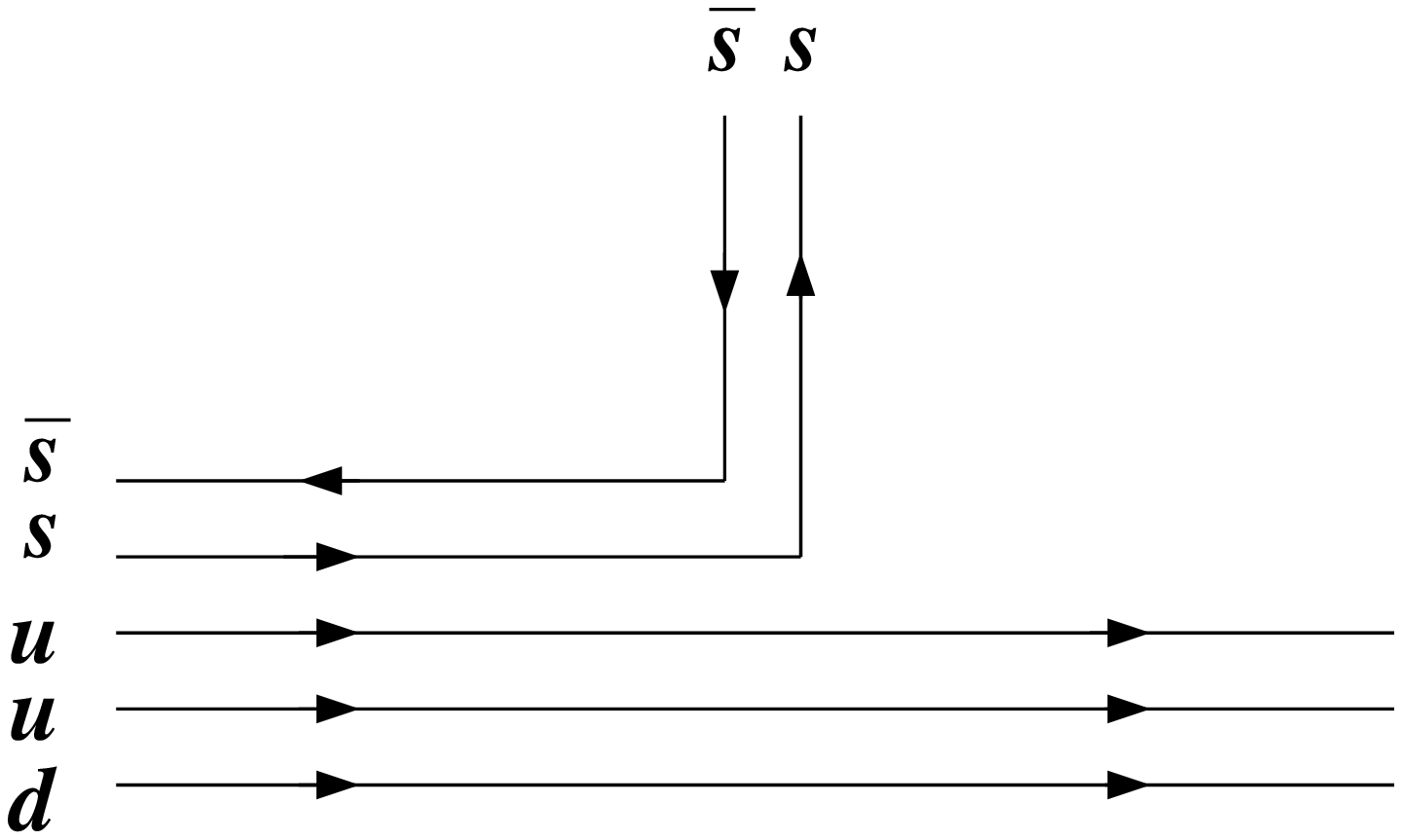}
\end{center}
\caption[]{\label{ssbar_hidden}
$\bar ss $ production from the hidden strangeness component of the proton.
{}For other notations, see text. 
(Gluon insertions are suppressed in this and the next two figures.)}
\end{figure}

\begin{figure}[htbp]
\begin{center}
\includegraphics*[width=6cm]{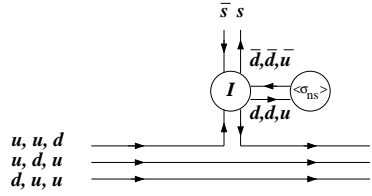}
\end{center}
\caption[]{\label{ssbar_prop}
$\bar ss $ production from the valence quarks in the proton due to
mass terms generation by the instanton-induced quark interaction coupled to 
spontaneous breaking of chiral symmetry.}
\end{figure}

\begin{figure}[htbp]
\begin{center}
\includegraphics*[width=7.5cm]{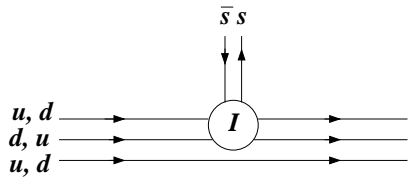}
\end{center}
\caption[]{\label{ssbar_direct}
Instanton-induced {\em direct} coupling of
$\bar s s$ to the nucleon.}
\end{figure}

\begin{figure}[htbp]
\begin{center}
\includegraphics*[width=7.5cm]{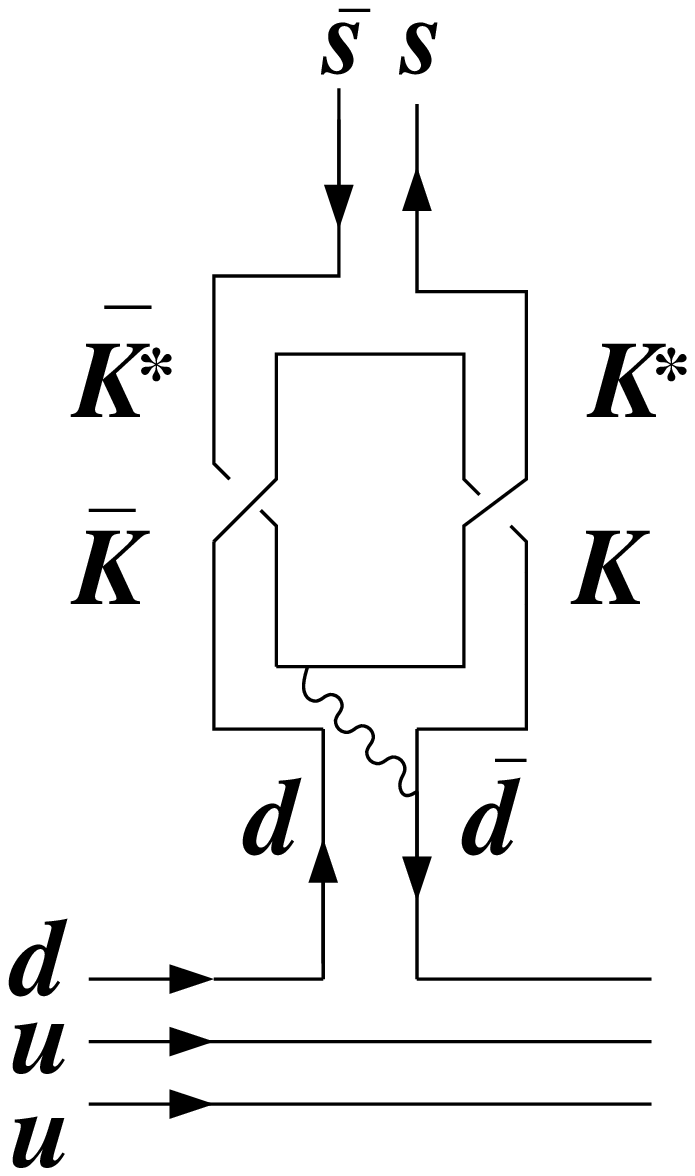}
\end{center}
\caption[]{\label{ssbar_loop}
Non-planar diagram
mechanism for the conversion of non-strange to strange 
quarkonium.}
\end{figure}

The co-existence of strange and non-strange quarkonia in the wave function of
the $\eta $ meson raises the question on the $\eta_{\rm{s}}$ creation by the
non-strange nucleon~\footnote{In this chapter we use the conventional notation
  $\eta_{\rm ns}$ and $\eta_{\rm s}$ for the non-strange and strange
  pseudoscalar fields instead of the notation $P_{\rm ns}$ and $P_{\rm s}$ of
  Section~\ref{eightfold}.}.  If the OZI rule were unbroken in the pseudoscalar sector,
the valence quarks of the non-strange nucleon could not contribute at all.
The only possible direct source for the $\eta_{\rm{s}} NN$ vertex would be the
hidden strangeness of the nucleon, i.e.  the existence of small, but
non--negligible $(uud)(\bar s s)$ configurations in the proton wave function,
see Fig.\,\ref{ssbar_hidden}.  The main $\bar s s $--source, however, is the
conversion of a non-strange quarkonium, emitted by the valence quarks (of the
nucleon), into the strange quarkonium under the influence of the OZI-rule
violating instanton effects. There seem to exist two mechanisms contributing
to this conversion. The first mechanism is displayed in Fig.\,\ref{ssbar_prop}
and has to do with mass terms generated by the anomaly when spontaneous
breaking of chiral symmetry takes place. In this case a $\bar qq$ pair is
replaced by its v.e.v and the flavor eigenstates $\eta_{\rm{s}}$ and 
$\eta_{\rm{ns}}$ 
get mixed -- an effect which has been quantified in Section~\ref{eightfold} in
the context of the broken linear sigma model.  The second mechanism, displayed
in Fig.\,\ref{ssbar_direct}, involves a {\it direct} instanton-induced
interaction.  Finally, less important sources for ($\bar s s$) quarkonia are
non-planar kaon loop diagrams of the type presented in Fig.\,\ref{ssbar_loop}.

In addition to the $\eta_{\rm s}$ production mechanisms mentioned above there
exist also contributions from $\eta_{\rm ns}$ production to the $\eta NN$
interaction. These contain the conventional mechanisms with continuous quark
lines in addition to instanton-induced interactions involving the hidden
strangeness of the nucleon. Clearly, the description of the $\eta NN$
interaction requires the disentanglement and quantification of all these
mechanisms. This can be accomplished in a natural way in the framework of the
model discussed in Section~\ref{eightfold}. 
To this end, we need to study the production
of flavor fields from the flavor axial currents which we discuss in the
framework of the same model.

The strange and non-strange weak decay constants are defined in the usual way:
\begin{equation}
 \langle 0|A^{\rm{ns}}_{\mu}(0)|\eta_{\rm{ns}}(q)\rangle
 =if_{\eta_{\rm{ns}} }q_\mu \;, \qquad  
 \langle 0|A^{\rm{s}}_{\mu}(0)|\eta_{\rm{s}}(q)\rangle
 =if_{\eta_{\rm{s}} }q_{\mu}\;  .  \label{wdc}
\end{equation}
Under the axial transformations
\begin{equation}
 \delta_A M = -(i/\sqrt{2})\{\epsilon^A ,M\} \; , \qquad
 \delta_A M^\dagger = (i/\sqrt{2})\{\epsilon^A ,M^\dagger \}
\end{equation}
with $\epsilon^A=\frac{1}{\sqrt{2}} \lambda_i \epsilon^A_i$,
the lagrangian in Eq.\,(\ref{L-model}) is not any longer invariant 
because of the explicit and instanton-induced 
symmetry breaking terms. 
A calculation of the divergences of the strange and 
non-strange  axial currents in the model yields:
\begin{equation}
 \partial^\mu A^{\rm{ns}}_{\mu}  = c_{\rm{ns}} \eta_{\rm ns} + 2\beta W\; ,
 \qquad \partial^\mu A^{\rm{s}}_{\mu}  =\sqrt{2} c_{\rm s} \eta_{\rm s} + 
 \sqrt{2}\beta W \; ,\label{divcurr}
\end{equation}
where W stands for the contribution of the instantons and contains 
trilinear, bilinear and linear terms in the fields. Explicitly
\begin{eqnarray}
 W&=& i( \det(M)\mbox{$-$}\det M^{\dagger} )\nonumber \\
&=& -2\sqrt{2}ab\eta_{\rm ns}
      -2a^2\eta_{\rm s} + bilinear   
   + trilinear\;. \label{W}
\end{eqnarray}
Taking the derivative of both  Eqs.\,(\ref{wdc}) and then inserting
(\ref{divcurr}) and 
(\ref{W}), we obtain
\begin{equation}
 f_{\eta_{\rm ns}} m^2_{\eta_{\rm ns}}= c_{\rm ns} 
 -4\sqrt{2} \beta ab\;, \qquad 
 f_{\eta_{\rm s}} m^2_{\eta_{\rm s}}
 =\sqrt{2} c_{\rm s}- 2\sqrt{2}\beta a^2 \;. \label{npcac}
\end{equation}
Notice that in the case when the anomaly is absent, the masses of both the
strange and non-strange fields (which in this fictitious case are the physical
fields) are solely 
driven by the explicit breaking terms (quark mass terms). Therefore these
fields are genuine pseudo-Goldstone boson fields in this case.  This property
is spoiled by the anomaly for both fields as can also be seen from the mass
relations
\begin{equation}
 m^2_{\eta_{\rm ns}}-m^2_\pi = -4\beta b\; , 
 \qquad 
 b m^2_{\eta_{\rm s}}+2\beta a^2= c_{\rm s}\;. 
 \label{m2-extra}
\end{equation}
We will come back to this point below. But first let us
insert the relations (\ref{m2-extra}) in Eq.\,(\ref{npcac}), such
that the non-strange and strange 
weak decay constants are determined as
\begin{equation}
 f_{\eta_{\rm ns}} =\sqrt{2}a=f_\pi \; , \qquad 
 f_{\eta_{\rm s}} = \sqrt{2} b = 2f_K -f_\pi\;.
 \label{f_eta_ns/s}
\end{equation}
These predictions  (see Table~\ref{input} for the results 
of models 1--3b) cannot not  be directly compared with the values 
$f_{\rm q}= (99 \pm 2)\,{\rm MeV}$ and 
$f_{\rm s}= (124 \pm 6)\,{\rm MeV}$ from Ref.~\cite{Feldmann2},
since the definitions are different, especially for $f_{\rm s}$ and
our $f_{\eta_{\rm s}}$.

In order to clarify whether the (octet-) eta is a 
pseudo-Goldstone boson or not,
let us discuss the fictitious case 
that the explicit breaking (\ref{c-equ-set})
in the non-strange sector, $c_{\rm ns}$, is put by hand to zero, 
but the one in the strange sector is kept finite, $c_{\rm s} \neq 0$.
Then only the pion remains a Goldstone-boson.
The kaon, however,
behaves as a pseudo-Goldstone boson, since in this case
its squared mass is proportional to the explicit
breaking $c_{\rm s} \propto m_{s}$.
Moreover, the non-vanishing of the (squared) $K$-meson mass
in the $m_\pi^2\to 0$ limit,
$m^2_K = (b-a)(\lambda b - 2 \beta ) \neq 0$, requires  that
$b\neq a$, i.e.\ that the v.e.v.\ generated by the
{\em spontaneous} breaking in the strange sector
differs from the one in the two non-strange sectors. For
this case it can be shown from (\ref{snsmixing1}--\ref{snsmixing3})
and (\ref{trace}--\ref{m2m2})  
that the physical $\eta$ meson
has the following squared mass
\begin{eqnarray}
m^2_{\eta} &=& -3 \beta b +{1\over 2}\lambda (b^2-a^2)
 -
\sqrt{(8a^2\mbox{+}b^2)\beta^2 + \lambda b (b^2\mbox{$-$}a^2) \beta +
{1\over 4}\lambda^2 (b^2\mbox{$-$}a^2)^2}\nonumber\\
& \neq& 0\; .
\label{meta_squared}
\end{eqnarray}
It is strongly affected by the anomaly since it
does not vanish in the case  $b\neq a$
unless the anomaly is nullified ($\beta=0$).
In this sense, the $\eta$ meson is not a Goldstone boson. But even
for finite $\beta<0$~\footnote{Remember that $\beta$ has to
be negative semi-definite as otherwise the squared $\eta$ and $\eta'$ masses
are not positive semi-definite.}, its mass still vanishes together with
the mass of the kaon, when $b\to a$. Because the v.e.v.\ $a$ 
does not need to vanish here, chiral
symmetry is spontaneously broken and the $\eta$ meson
is then a {\em pseudo\/}--Goldstone boson, although strongly affected by 
the anomaly. Note that the mass of the $\eta'$ meson does not vanish
in the last case and therefore the $\eta'$ cannot  become a pseudo-Goldstone
boson for a non-vanishing (negative) $\beta$ and $b\to a\neq 0$, 
since in the expression of $m_{\eta'}^2$ the analogous 
square root to the one in Eq.\,(\ref{meta_squared})
shows the opposite sign.

As a starting point in the analysis of the $\eta NN$ coupling we will assume
that the anomaly is absent, i.e.\ the parameter $\beta=0$. 
In this case the $\eta_{\rm ns}$ couples to the
non-strange nucleon only via the $u,d$ (valence and sea) components of the
nucleon, whereas the $\eta_{\rm s}$-nucleon coupling is solely induced by the
hidden $\bar s s$ component of the nucleon. The $\eta_{\rm{s}}$ and 
$\eta_{\rm{ns}}$ fields are genuine Goldstone bosons in this case, and 
we can assume -- without loss of generality -- that their
interaction with the nucleon follows the derivative structure
(as it has to be the case in the non-linear realization
of the pseudoscalar fields in chiral perturbation theory)
\begin{eqnarray}
 {\mathcal L}_{\eta_{\rm ns} NN} &=&
 \left({\partial_\mu \eta_{\rm ns} \over f_{\eta_{\rm ns}}}\right)\, 
 \,g_A^{\rm ns} \bar \Psi _N \gamma^\mu \gamma^5{1\over 2}  \Psi_N \;,
  \label{chiral_etansNN}
 \\
 {\mathcal L}_{\eta_{\rm s} NN} &=& 
 \left({\partial_\mu \eta_{\rm s}\over f_{\eta_{\rm s}}}\,\right) 
 \,g_A^{\rm s} \bar \Psi _N \gamma^\mu \gamma^5{1\over 2}  \Psi_N \;,
  \label{chiral_etasNN}
\end{eqnarray}
where $g^{\rm ns}_A = a_u+a_d$ and $g^{\rm s}_A = \sqrt{2} a_s$.
The quantities $a_u$, $a_d$, and $a_s$ are the
fractions of proton spin carried by the $u$, $d$, and $s$ quark seas
(including valence contributions),
respectively, and are known from deep inelastic scattering data
\cite{proton}. They are parameterized as
\begin{eqnarray}
 a_{q_i}(Q^2) &=& \Delta q_i(Q^2) -
 {\alpha_s(Q^2)\over {2\pi }}\Delta g(Q^2) \; ,
 \quad q_i =u,d,s\;,\nonumber\\
 Q^2 &=& -(p-p')^2\; ,
 \label{spin_fr}
\end{eqnarray}
where $\Delta q_i (Q^2)$ is the genuine spin fraction associated 
with the $q_i$ flavored quark sea, while the $\Delta g (Q^2)$  
term describes de-polarization effects due to gluon contributions.
We use below  the values of the spin-fractions reported in 
\cite{proton} for $Q^2=10\,{\rm GeV}^2$ as $a_u =0.82\pm 0.02$,
$a_d=-0.44 \pm 0.02 $, and $a_s=-0.10\pm 0.02$.

In Eqs.\,(\ref{chiral_etansNN},\ref{chiral_etasNN}) it is {\em assumed} that
the $\eta_{\rm ns}$ and $\eta_{\rm s}$ are the flavor-eigenstates of a
pseudo-vectorial $\eta$-nucleon interaction. For this reason, the interactions
(\ref{chiral_etansNN}) and (\ref{chiral_etasNN}) are designed as the
non-strange and strange equivalents of the standard $\pi N$ Goldberger-Treiman
coupling, respectively.  In the spirit of an effective chiral expansion,
${\mathcal L}_{\eta_{\rm ns} NN}$ and ${\mathcal L}_{\eta_{\rm s}NN}$ would 
give the {\em leading} axial couplings in these two sectors.  The latter
would be a sensible approximation if $\eta_{\rm ns}$ and $\eta_{\rm s}$ could
be treated as Goldstone bosons.  However, as our discussion in 
Section~\ref{eightfold}
showed, both states are strongly affected by the anomaly, and therefore we
must be careful about this point.

Let us first focus on the effects due to the (anomaly-induced) mixing
of flavor fields during propagation. The calculation of these contributions to
the $\eta NN$ coupling can be formulated in a coupled-channel scheme.  In this
formalism the eta-state can be written as a vector
in the $\eta_{\rm ns}$ -- $\eta_{\rm s}$ space: 
\begin{equation}
 \left( \begin{array}{c} \cos \phi_P  \\ -\sin \phi_P 
 \end{array} \right) 
 \label{vector_space}
\end{equation}
(see Eq.\,(\ref{etawafu})).
In the case when the anomaly is turned off, this state couples (in its 
transposed form) 
via the diagonal `off-shell' propagator to  the axial vector  
current of the nucleon, i.e.\
\begin{equation}
 \label{prop}
 \left(\cos \phi_P  , - \sin\phi_p\right) \times 
 \left( \begin{array}{cc}  \frac{i}{q^2- m^2_{\eta_{\rm ns}}} & 0 \\
     0 & \frac{i}{q^2- m^2_{\eta_{\rm s}}} \end{array}
 \right) \times \left( \begin{array}{c}\frac{g_A^{\rm ns}}
 {f_{\eta_{\rm ns}}}\\
 \frac{g_A^{\rm s} }{f_{\eta_{\rm s}}} \end{array} 
  \right)\, i q_\mu  \bar \Psi_N \gamma^\mu \gamma^5 \frac{1}{2} \Psi_N \; .
\end{equation}
In other words, the
strange--non-strange fields are the propagating fields.

Let us now consider the physical case when the anomaly is present. 
As a consequence, one encounters
the mixing of the flavor fields 
{\it during propagation\/} rendering propagating $\eta$ and $\eta'$. 
For this reason, one has to replace
the diagonal propagator matrix in Eq.\,(\ref{prop})  by
the full propagator matrix including the anomaly-induced non-diagonal 
elements\,:
\begin{eqnarray} 
 i \left( \begin{array}{rr} q^2\mbox{$-$}m_{\eta_{\rm ns}}^2   & \quad 
 -m_{\eta_{\rm s-ns}}^2 \\
                   -m_{\eta_{\rm s-ns}}^2 & \quad q^2\mbox{$-$}
m_{\eta_{\rm s}}^2 
         \end{array} \right)^{-1} \!\!\!
 =
 \frac{i}{(q^2\mbox{$-$}m^2_\eta )(q^2\mbox{$-$}m^2_{\eta'})}
 \left( \begin{array}{rr}q^2\mbox{$-$}m_{\eta_{\rm s}}^2  
 & \quad m_{\eta_{\rm s-ns}}^2 \\
                            m_{\eta_{\rm s-ns}}^2 
 & \quad q^2\mbox{$-$}m_{\eta_{\rm ns}}^2 
         \end{array} \right)\; . \nonumber\\
\label{fullprop}
\end{eqnarray} 
With the aid of the following relations which arise from the 
diagonalization of the pseudoscalar isoscalar sector
\begin{eqnarray} \nonumber
 m^2_{\eta_{\rm s}}\cos\phi_P   + m^2_{\eta_{\rm s-ns}} \sin\phi_P  &=& 
 m^2_{\eta'}\cos\phi_P  \;,\\ \nonumber
 m^2_{\eta_{\rm ns}}\sin\phi_P   + m^2_{\eta_{\rm s-ns}} \cos\phi_P  &=& 
 m^2_{\eta'}\sin\phi_P  \;, \\ \nonumber
 m^2_{\eta_{\rm s}}\sin\phi_P   - m^2_{\eta_{\rm s-ns}} \cos\phi_P  &=& 
 m^2_{\eta}\sin\phi_P  \;,\\ 
 m^2_{\eta_{\rm ns}}\cos\phi_P   - m^2_{\eta_{\rm s-ns}} \sin\phi_P  &=& 
 m^2_{\eta}\cos\phi_P  \;, \label{mixrel}
\end{eqnarray}
one can cast Eq.\,(\ref{prop}), 
where the diagonal propagator matrix is now replaced by the
full propagator matrix 
Eq.\,(\ref{fullprop}), into the following form:
\begin{equation}
 \frac {i}{q^2-m^2_{\eta}} \left( \frac {g^{\rm ns}_A}{f_{\eta_{\rm ns}}}
 \cos \phi_P  
 -  \frac {g^{\rm s}_A}{f_{\eta_{\rm s}}} \sin \phi_P  \right)  
 i q_{\mu}\bar \Psi _N \gamma^\mu \gamma^5{1\over 2}  \Psi_N\; .
\end{equation}
This last equation illustrates 
the formation of propagating $\eta$ and (by a similar calculation) $\eta'$
fields under the influence of anomaly-induced mixing effects 
{\em during the propagating stage} of 
the isoscalar pseudoscalars.
The coupling strengths of the $\eta$ and $\eta'$ fields
to the nucleon are determined as 
\begin{equation}
 g_{\eta NN} =  \frac {g^{\rm ns}_A M_N}{f_{\eta_{\rm ns}}} \cos \phi_P  
 -  \frac {g^{\rm s}_A M_N}{f_{\eta_{\rm s}}} \sin \phi_P  \label{getaNN}
\end{equation}
and 
\begin{equation}
 g_{\eta_\prime NN} 
 =  \frac {g^{\rm ns}_A M_N}{f_{\eta_{\rm ns}}} \sin \phi_P  
 + \frac {g^{\rm s}_A M_N}{f_{\eta_{\rm s}}} \cos \phi_P \;,\label{getapNN}
\end{equation}
respectively.
Eqs.\,(\ref{getaNN},\ref{getapNN}) become exactly the octet and singlet 
relations for the case that $\phi_P =\phi_{\rm su(3)}$ 
(i.e.\ $\cos\phi_P  = \sqrt{1/3}$ and $\sin\phi_P = \sqrt{2/3}$) and that 
$f_{\eta_{\rm ns}}$ = $f_{\eta_8}$ = $f_{\eta_{\rm s}}$ = $f_{\eta_0}$
= $f_K$ = $f_\pi $:
\begin{equation}
 g_{\eta NN} =  \sqrt{\frac{1}{3}} \,\frac{M_N}{f_\pi} a_8 \;, \qquad
 g_{\eta' NN} =  \sqrt{\frac{2}{3}} \,\frac{M_N}{f_\pi} a_0 \; ,
 \label{naive-oct-sing}
\end{equation}
where $a_8=a_u+a_d-2a_{\rm s},~ a_0=a_u+a_d+a_{\rm s}\,$.

The expressions (\ref{naive-oct-sing}) 
for $g_{\eta NN}$ and $g_{\eta'NN}$ are the well
known octet- and the singlet Goldberger-Treiman relations, respectively.
In the literature, as discussed in Section 3.11 of Ref.\,\cite{Feldmann2}, 
there exists  
an other alternative for the description of 
the Goldberger-Treiman relation in the singlet channel.
Shore and Veneziano~\cite{ShoVe} established
a two-component description  of the singlet axial charge
where the singlet Goldberger-Treiman relation is modified by
an additional direct coupling of the pseudoscalar 
operator $G \widetilde G$ 
to the nucleon,
\begin{eqnarray}
\sqrt{\frac{2}{3}}\,\frac{ M_N}{\widetilde f} a_0 =  g_{\eta' NN} + 
\sqrt{\frac{2}{3}} \,
{\widetilde f} m^2_{\eta'} g_{\widetilde G NN} . 
\end{eqnarray}
Here  $G$ stands for the gluon field, $\widetilde{G}$ for  its dual,
and the new couplings 
$g_{\eta' NN}$  and $g_{\widetilde G NN}$ are
related to the polarized quark distributions and
polarized gluon distribution, respectively.
As it is stressed in Ref.\,\cite{Feldmann2}, the quantity
$\widetilde f$ does not coincide with
the decay constant  $f_{\eta_0}$. The latter scales as $a_0$,
whereas $\widetilde f$ is scale-independent.
We refer the reader to  Ref.\,\cite{Feldmann2} for
further details and a comparison of both schemes.


The above considerations could be criticized on two accounts~\footnote{Of 
course, the omission of  direct instanton-induced couplings of the type
displayed  in
Fig.\,\ref{ssbar_direct} and the neglect of subleading non-planar 
$\bar K^{(\ast)} K^{(\ast)}$ loops, see Fig.\,\ref{ssbar_loop},
are further points open to criticism, but outside the scope of the model.}.
First, they are based on special choice of couplings of the
flavor-eigenstates $\eta_{\rm ns}$ and $\eta_{\rm s}$ to the nucleon.
Secondly, the role of the anomaly in the $\eta$, $\eta'$ production
at the $\eta,\eta'$-nucleon vertex
is not explicit.
Below, we will close these gaps in the derivation by adapting
the usual derivation of the
Goldberger-Treiman relation for pions to the $A^\mu_{\rm ns}$ and
$A^\mu_{\rm s}$ cases and keeping track of the 
modifications caused by the $U_A(1)$ anomaly. The
divergences of these currents (see Eq.\,(\ref{divcurr})), 
when sandwiched between nucleon states,
clearly exhibit some of
the anomaly effects in the $\eta (\eta')NN$ interaction.

On the basis of symmetries (Lorentz covariance, parity etc.), the matrix 
elements of the flavor currents between nucleon states can be 
parameterized as follows
\begin{eqnarray}\nonumber
 \langle N'|A^{\rm ns}_\mu |N\rangle 
 &=& \bar u (p',s')\left[\gamma_\mu\gamma_5 G^{\rm ns}_A
 (q^2) + q_\mu\gamma_5 G^{\rm ns}_P(q^2)\right]\frac{1}{2}u(p,s)\;, \\
 \langle N'|A^{\rm s}_\mu |N\rangle 
 &=& \bar u (p',s')\left[\gamma_\mu\gamma_5 G^{\rm s}_A
 (q^2) + q_\mu\gamma_5 G^{\rm s}_P(q^2)\right]\frac{1}{2}u(p,s)\;,
\end{eqnarray}  
where $q=(p'-p)$ is the transferred momentum, $s$ and $s'$ denote nucleon
polarizations, and $G^{\rm ns,s}_{A}(q^2)$ and $G^{\rm ns,s}_{P}(q^2)$ are the
axial vector and induced pseudoscalar form factors, respectively. These form
factors can not be fixed on symmetry grounds alone. The divergence of
the above matrix elements yields
\begin{eqnarray}\label{parac}\nonumber
 \langle N'|i\partial^\mu A^{\rm ns}_\mu |N\rangle &=& 
 -\bar u (p',s')\left[2 M_N G^{\rm ns}_A
 (q^2) + q^2 G^{\rm ns}_P(q^2)\right]\frac{\gamma_5}{2}u(p,s)\\ \nonumber
 &=&i\langle N'|{\rm c_{ns}}\eta_{\rm ns} + 2 \beta W |N\rangle , \\
 \langle N'|i\partial^\mu A^{\rm s}_\mu |N\rangle 
 &=& -\bar u (p',s')\left[2 M_N G^{\rm s}_A
 (q^2) + q^2 G^{\rm s}_P(q^2)\right]\frac{\gamma_5}{2}u(p,s)\\ \nonumber
 &=&i\langle N'|\sqrt{2}
 {\rm c_ s}\eta_{\rm s} + \sqrt{2} \beta W |N\rangle\;.  
\end{eqnarray} 
Here use has been made of Eqs.\,(\ref{divcurr}). In combining
Eqs.\,(\ref{W}) and (\ref{npcac}) one arrives at
\begin{eqnarray} \nonumber
 {\rm c_{\rm ns}}\eta_{\rm ns} + 2 \beta W 
 =f_{\eta_{\rm ns}}m^2_{\eta_{\rm ns}}\eta_{\rm ns} 
 -4\beta a^2 \eta_{\rm s} + bilinear + trilinear\;, \\
 \sqrt{2}{\rm c_s}\eta_{\rm s} + \sqrt{2} \beta W 
 =f_{\eta_{\rm s} }m^2_{\eta_{\rm s}}
 \eta_{\rm s} -4\beta ab \eta_{\rm ns} + bilinear + trilinear\;. 
 \label{simplif}
\end{eqnarray}
At low energies, the matrix elements of the  pseudoscalar flavor 
fields between nucleons are dominated by the exchange of propagating 
$\eta$ and $\eta'$ mesons
\begin{eqnarray}\nonumber
 i\langle N'|\eta_{\rm ns}|N\rangle&=&\left( \frac{g_{\eta NN}}{q^2-m^2_\eta }
 \langle\eta_{\rm ns}|\eta\rangle + \frac{g_{\eta' NN}}{q^2 - m^2_{\eta'} } 
 \langle\eta_{\rm ns}|\eta' \rangle \right) \bar u(p',s')\gamma_5 u(p,s)\;, \\
 i\langle N'|\eta_{\rm s}|N\rangle&=&\left( \frac{g_{\eta NN}}{q^2-m^2_\eta }
 \langle\eta_{\rm s}|\eta\rangle + \frac{g_{\eta' NN}}{q^2-m^2_{\eta'}} 
 \langle\eta_{\rm s}|\eta' \rangle \right) \bar u(p',s')\gamma_5 u(p,s)
 \;.
 \label{flavpNN} 
\end{eqnarray}
In deriving Eqs.\,(\ref{flavpNN}) we used the following $\eta (\eta')NN$ 
lagrangian 
\begin{equation}
 {\mathcal L}= g_{\eta NN}\, \eta \,
 \bar\Psi_N i\gamma_5 \Psi_N +  g_{\eta' NN} \,\eta'\, 
 \bar\Psi_N i\gamma_5 \Psi_N\;.  
\end{equation}
Note that we now apply  couplings of pseudoscalar nature, in contrast
to the former derivation which was  based on couplings of derivative 
nature, see  Eqs.\,(\ref{chiral_etansNN}) 
and (\ref{chiral_etasNN}).
Inserting (\ref{simplif},\ref{flavpNN}) in Eqs.\,(\ref{parac}) and 
disregarding  contributions from bilinear and trilinear terms in the 
anomaly leads to 
\begin{eqnarray}\nonumber
 \frac{1}{2}G^{\rm ns}_P (q^2)&=&\frac{-M_N G^{\rm ns}_A(q^2)
 \mbox{+}g_{\eta NN}
 F^{\rm ns}_\eta \mbox{+}g_{\eta' NN} F^{\rm ns}_{\eta'}}{q^2}
-\frac{g_{\eta NN}
 F^{\rm ns}_\eta }{q^2-m^2_\eta} - \frac{g_{\eta' NN}
 F^{\rm ns}_{\eta'} }{q^2-m^2_{\eta'}}\;, \\
 \frac{1}{2}G^{\rm s}_P(q^2)&=&\frac{-M_N G^{\rm s}_A(q^2)\mbox{+}g_{\eta NN}
 F^{\rm s}_\eta \mbox{+}g_{\eta' NN}F^{\rm s}_{\eta'}}{q^2}-\frac{g_{\eta NN}
 F^{\rm s}_\eta }{q^2-m^2_\eta} - \frac{g_{\eta' NN}
 F^{\rm s}_{\eta'} }{q^2-m^2_{\eta'}} \;. \nonumber\\
\label{F2F1}
\end{eqnarray}
Here, we used the definitions
\begin{equation}
\begin{array}{lclcr}
 F^{\rm ns}_\eta &\equiv &\frac{1}{m^2_{\eta}}
 \left( f_{\eta_{\rm ns}}m^2_{\eta_{\rm ns}}\cos 
 \phi_P  + 4\beta a^2 \sin \phi_P   \right) &=& f_{\eta_{\rm ns}} \cos \phi_P
\;,\\
 F^{\rm ns}_{\eta'}&\equiv& \frac{1}{m^2_{\eta'}}
 \left( f_{\eta_{\rm ns}}m^2_{\eta_{\rm ns}}
 \sin \phi_P   - 4\beta a^2 \cos \phi_P   \right)
&=& f_{\eta_{\rm ns}} \sin \phi_P
\;, \\  
 F^{\rm s}_\eta &\equiv &\frac{1}{m^2_{\eta}}
 \left( -f_{\eta_{\rm s}}m^2_{\eta_{\rm s}}\sin 
 \phi_P   - 4\beta ab \cos \phi_P   \right)
&=& - f_{\eta_{\rm s}} \sin\phi_P 
\;,\\
 F^{\rm s}_{\eta'}&\equiv& \frac{1}{m^2_{\eta'}}
 \left( f_{\eta_{\rm s}}m^2_{\eta_{\rm s}}\cos 
 \phi_P   - 4\beta ab \sin \phi_P   \right)
&=& f_{\eta_{\rm s}} \cos \phi_P \;,
\end{array}
\label{Rfp} 
\end{equation}
where the r.h.s.\ 
equations follow from the relations (\ref{mixrel}),
the expressions (\ref{f_eta_ns/s}) of $f_{\eta_{\rm ns}}$ and 
$f_{\eta_{\rm s}}$ and $m^2_{\eta_{\rm s-ns}}$ as defined in 
 (\ref{snsmixing3}).
Thus the
$ F^{\rm ns,s}_{\eta,\eta'}$ are nothing but
the weak decay constants for 
the production of physical $\eta$, $\eta'$ fields by 
the  flavor currents (see Eqs.\,(\ref{wdc})),
i.e.
\begin{equation}
\begin{array}{lclclcl}
\langle 0|A^{\rm ns}_{\mu}|\eta (q)\rangle  &=& i F^{\rm ns}_{\eta}
q_{\mu}\; , & \qquad &
\langle 0|A^{\rm ns}_{\mu}|\eta' (q)\rangle  &=& i F^{\rm ns}_{\eta'}
q_{\mu}\; ,\\
\langle 0|A^{\rm s}_{\mu}|\eta (q)\rangle  &=& i F^{\rm s}_{\eta}
q_{\mu}\; , & \qquad &
\langle 0|A^{\rm s}_{\mu}|\eta' (q)\rangle  &=& i F^{\rm s}_{\eta'}
q_{\mu}\; ,\\
\end{array}
\end{equation} 
such that  the isoscalar pseudoscalar versions of the PCAC relations 
are implied (see Eqs.\,(\ref{divcurr})
and (\ref{npcac})):
\begin{equation}
\begin{array}{lclclcl}
\langle 0|\partial^\mu A^{\rm ns}_{\mu}|\eta (q)\rangle  &=& 
F^{\rm ns}_{\eta} m_\eta^2\; , & \qquad &
\langle 0|\partial^\mu A^{\rm ns}_{\mu}|\eta' (q)\rangle 
 &=&  F^{\rm ns}_{\eta'} m_{\eta'}^2\;\; ,\\
\langle 0|\partial^\mu A^{\rm s}_{\mu}|\eta (q)\rangle 
 &=&  F^{\rm s}_{\eta} m_\eta^2\; , & \qquad &
\langle 0|\partial^\mu A^{\rm s}_{\mu}|\eta' (q)\rangle  
&=&  F^{\rm s}_{\eta'} m_{\eta'}^2 \;.\\
\end{array}
\end{equation} 
The poles at $q^2=0$ in the $G^{\rm ns,s}_P(q^2)$ form factors in 
Eq.\,(\ref{F2F1}) are unphysical 
since in the explicitly broken case 
there exist no massless excitation. These terms can 
be eliminated, if one requires that the corresponding numerators vanish. 
This condition 
leads to the following 
relations 
\begin{equation}
\begin{array}{lcr}
 M_N G^{\rm ns}_A(q^2)&=& g_{\eta NN}f_{\eta_{\rm ns}}\cos\phi_P 
 + g_{\eta' NN} 
 f_{\eta_{\rm ns}}\sin \phi_P\;, \\
 M_N G^{\rm s}_A(q^2)&=& 
-g_{\eta NN} f_{\eta_{\rm s}}\sin\phi_P + g_{\eta' NN} 
 f_{\eta_{\rm s}}\cos\phi_P\;.
\end{array}
 \label{eepGTR}
\end{equation}
Eqs.\,(\ref{eepGTR}) are the $\eta,\eta'$ analogs of the Goldberger-Treiman
relation for pions~\footnote{As a cross-check notice that in the case of
  vanishing $\beta$, the mixing angle $\phi_P $ is zero and 
  Eqs.\,(\ref{eepGTR}) become simply the Goldberger-Treiman relations of the
  generation-flavor fields as postulated in
  Eqs.\,(\ref{chiral_etansNN},\ref{chiral_etasNN}).}.  
They are valid for
small $q^2$ and under the condition that the $G^{\rm ns,s}_A$ form factors
change slowly with $q^2$ (they have no poles) in this energy region.  One can
easily check that Eqs.\,(\ref{eepGTR}) are compatible with the relations
(\ref{getaNN}) and (\ref{getapNN}), if at $q^2=0$ the axial vector form
factors are expressed by the spin fractions $ G^{\rm ns}_A (0)=g^{\rm ns}_A =
a_u + a_d, ~~ G^{\rm s}_A (0)=g^{\rm s}_A = \sqrt{2} a_s$.  Thus both
derivations are completely consistent: the here discussed contributions of the
anomaly to the creation of physical pseudoscalar fields 
from the vacuum are hidden in the
$\sin\phi_P$ and $\cos\phi_P$ terms (\ref{vector_space}) 
in the coupled channel scheme.  There is
no ``direct instanton induced interaction'' in this calculation.  The place to
include such a contribution are the $G^{\rm s,ns}_A (q^2)$ form factors which
here were solely identified with the spin fractions.  

The conclusion we
extract in this model-dependent analysis of the non-per\-turbative effects
brought about by 't Hooft's effective  
instanton-induced interaction, is the following:
this interaction, in addition to being responsible for the $\eta $ as a member
of the pseudoscalar octet, is also responsible for its PCAC behavior and
the interactions of the $\eta$
meson with external sources.  These interactions exhibit an octet-like
behavior in the case we consider instanton effects during propagation or
creation from the vacuum of pseudoscalar fields. Our expectation is that
deviations from this behavior are related to the direct instanton induced
interactions shown in Fig.\,\ref{ssbar_direct} and to the inclusion
of subleading non-planar $\bar K^{(\ast)} K^{(\ast)}$ loops.

Using the results and input-values of the models of Table~\ref{input} for the
quantities involved and the spin-fraction $a_{q_i}$ as defined below
Eq.\,(\ref{spin_fr}), we obtain from the re-derived 
Eqs.\,(\ref{getaNN}) and (\ref{getapNN})
\begin{eqnarray} 
 g_{\eta NN} &=&2.8 \pm 0.5 \quad\mbox{and}\quad g_{\eta'NN}= 2.8 \pm 0.4 
 \qquad\mbox{(model 1)}\;,\\
 g_{\eta NN} &=&3.3 \pm 0.6 \quad\mbox{and}\quad g_{\eta'NN}= 2.3 \pm 0.6
 \qquad\mbox{(model 2)}\;,\\
 g_{\eta NN} &=&3.5 \pm 0.7 \quad\mbox{and}\quad g_{\eta'NN}= 2.0 \pm 0.7
 \qquad\mbox{(model 3a)}\;, \\
 g_{\eta NN} &=&3.9 \pm 0.7 \quad\mbox{and}\quad g_{\eta'NN}= 1.2 \pm 0.7
 \qquad\mbox{(model 3b)}\;, 
\end{eqnarray}
where the errors of $g_{\eta NN}$ and $g_{\eta' NN}$ are partially correlated
because of the spin fractions and the very precisely determined value 
of  $g_A^3=a_u-a_d=1.267 \pm 0.004 $, 
and therefore
\begin{eqnarray} 
 \frac{g_{\eta NN}^2}{4\pi}+\frac{g_{\eta' NN}^2}{4\pi}&=& 1.2\pm 0.4 \qquad
 \mbox{(model 1)}\;,
 \label{combiI}\\
 \frac{g_{\eta NN}^2}{4\pi}+\frac{g_{\eta' NN}^2}{4\pi}&=& 1.3\pm 0.4 \qquad
 \mbox{(models 2, 3a, and 3b)}\;, \label{combiII}
\end{eqnarray}
see Table~\ref{input}.  
Note that the displayed
error bars have been calculated solely from the errors $\pm 0.02$ of the spin
fractions $a_{q_i}$ (which contribute about 70--80\,\% of the total error),
from the error $\pm 4\,{\rm MeV}$ of $f_{\pi^0}$\,\cite{PDG} (which
has to be used here instead of the more 
precisely determined $f_{\pi^\pm}$), and from other
uncertainties in the input quantities.  Further sources of (systematical) 
errors could result
from the extrapolation from the $q^2=0$ point of weak interaction to the
mass-shells $m^2_\eta$ and $m^2_{\eta'}$ of strong interactions, from higher
order corrections in the symmetry breaking, from meson-loop corrections, from
the neglect of the subleading OZI-rule-violating disconnected hairpin diagrams
(e.g., from an incomplete cancellation of the non-planar 
$\bar K^{(\ast)}  K^{(\ast)}$ loops), and from the
neglect of baryon resonances as e.g.\ the $S_{11}(1535)$.  The values of all
four models are still -- within the errors -- compatible with the upper bound
obtained by Grein and Kroll from the analysis of {\it NN} forward
scattering~\cite{GrKr}
\begin{equation}
 \frac{g_{\eta NN}^2}{4\pi}
 +\frac{g_{\eta' NN}^2}{4\pi} \stackrel{<}{\sim} 1 \; .
\end{equation}
Note that the high-precision measurements of the differential cross sections
in $\eta $ photo-production off--proton near threshold at the Mainz Microtron
(MAMI) \cite{MAMI} were interpreted in \cite{TiKa} in terms of a strongly
suppressed $g_{\eta NN}$ value of $|g_{\eta NN}|\approx 2.25\pm 0.15 $ (or,
equivalently $g_{\eta NN}^2 /4\pi \approx 0.4$).  This result was concluded on
the basis of the small $P$-wave contribution to the almost flat angular
distributions for a wide range of beam energies. Only our model 1 is
compatible with the value of Ref.\,\cite{TiKa}.  {}From a measurement of
$\eta'$ production in proton-proton collisions close to threshold at COSY
\cite{Moskal} a bound $|g_{\eta' NN}| \leq 2.5 $ has been deduced which is
compatible with the predictions of all our models within  error bars.

The significant suppression of the $\eta N$ coupling relative
the octet Goldberger-Treiman relation was noticed by several
authors, the most recent being, among others 
Refs.~\cite{KDO,Zhu,Saghai}.
In Ref.~\cite{KDO} $g^2_{\eta NN}/4\pi $ was evaluated 
from QCD sum rules beyond the chiral limit and predicted to
range from the still relevant value of
$g^2_{\eta NN}/4\pi =0.42$ (for the $SU(3)$ limit)
down to the almost complete decoupling of 
$g^2_{\eta NN}/4 \pi
=0.03$  (beyond the $SU(3)$
limit). 
Light cone QCD sum rules also lead to the small value of
$g_{\eta NN}^2/4\pi =0.3\pm 0.15$. Finally, the value
of $g_{\eta NN}^2/4\pi =0.1\pm 0.01$ was concluded on the basis
of  recent quark model analyses of meson-photo production data performed
in \cite{Saghai}. 
In the present work, we  have 
revealed the nature of the subtle mechanism of the 
formation of the octet flavor symmetry under the umbrella of the anomaly
as reflected by 't Hooft's effective instanton induced interaction.
Our results strongly hint onto the possibility that
the reasons for the observed suppression
of $g_{\eta NN}$ have to be searched for beyond  
't Hooft's effective interaction.

The naive octet-singlet scheme of Eq.\,(\ref{naive-oct-sing}) predicts
$g_{\eta NN} = 3.4\pm 0.6$ and $g_{\eta'NN}= 2.3\pm 0.6$ and $1.4 \pm 0.4$ for
the Grein-Kroll bound.  Note that these results 
more or less agree with the ones of  our model 2.
This can be justified from the fact that the mixing angle of this model
is the closest to the octet--singlet case, with the exception of model 1.
The latter, however, features a rather large value of $f_K$ which severely 
breaks
the condition $f_K \approx f_\pi$ specified above Eq.\,(\ref{naive-oct-sing}).
 Furthermore, the values $g_{\eta NN} = 3.4 \pm 0.5$ and
$g_{\eta' NN}= 1.4 \pm 1.1$ of Ref.\,\cite{Feldmann2} 
are calculated with a mixing
angle $\phi_P\approx 39.3^\circ$ which corresponds to $\theta_P \approx
-15^\circ$. Thus this result falls  
between the ones of model 3a and 3b which are
fixed by the mixing angles $\theta_P = -10^\circ$ and $\theta_P=-20^\circ$.
The rule is that the value of $g_{\eta NN}$ increases and the one of $g_{\eta'
  NN}$ decreases with decreasing mixing angle, while the
``Grein-Kroll-strength''  is nearly constant
(see Eqs.\,(\ref{combiI}) and (\ref{combiII})).

\section{Summary}
Admittedly, we have used a rather special model in order to analyze the
OZI-rule respecting basis of the meson sector and the consequences for the
$\eta,\eta'$-nucleon coupling constants, namely the chirally symmetric $U(3)_L
\otimes U(3)_R$ {\em linear} sigma model. Note, however, that the
model-dependence is manifest for the scalar sector, whereas the tree-level
predictions of the model in the pseudoscalar sector, under the
same input of course, ought to be compatible
with PCAC arguments or tree-level calculations in chiral perturbation theory,
if, and that is the important point, the modifications of the anomaly are
properly incorporated into the latter.  Furthermore, we have explicitly
assumed that the OZI-violation in the isoscalar-pseudoscalar meson sector is
governed to leading order by instanton-induced effects rather than by large
$N_c$ effects, i.e.
that meson loops such as the non-planar OZI-violating diagrams,
give a non-zero, but {\em subleading} contribution to this sector.  
These points are still
controversially discussed in the literature, see e.g.\ \cite{schshur} and
\cite{DeGrand} which
favor the instanton interpretation and criticize \cite{isgur} and
\cite{isgur2}, respectively (and vice versa),
where the
large $N_c$ interpretation (see also \cite{Feldmann2,KaiLeut}) and the
importance of the non-planar meson-loop contributions is advocated for.

However, if we take our assumptions as stated, we find that the mixing of
non-strange and strange quarkonia in the wave function of the physical $\eta $
meson induced by the determinantal instanton-induced 't~Hooft interaction is
such that the $\eta $ meson is close to the octet state.
Therefore, the unitary spin symmetry is more obliged for its
existence to the effects of the
the axial gluon anomaly than being a fundamental symmetry in its
own rights. 
The model also allows us to study consequences for the $\eta NN$ coupling. 
In two independent calculations, one based on the conventional derivative
coupling of the flavor-eigenstates to the nucleon, and the other based
on a careful study of the axial vector coupling {\em including anomaly
contributions} (resulting from the 't Hooft interaction) to the nucleon,  
we obtained the magnitude of $g_{\eta NN}$. We found it, within error
bars, to be
stable against changes in the input parameters and in addition, to be
be close to  the ordinary $SU(3)$ results.
Though we obtained our $g_{\eta NN}$ to obey a Goldberger-Treiman type
relation, the latter {\it did not\/} take its origin from
a (massless) pole dominance of the induced pseudoscalar form factor.
Rather it appeared as a consequence of the subtle effect
of instanton induced flavor-mixing {\it during propagation\/} of
the isoscalar pseudoscalars.

In having clarified the role of the axial gluon anomaly
(as mimicked by 't Hooft's effective instanton induced interaction)
for manufacturing  the octet way, we  have established the limits 
beyond which one has to extend the model in order
to describe possible deviations of the $g_{\eta NN}$ value from
its octet-GT-value.
 Among the possible
candidates for such effects we emphasize 
the direct meson-instanton coupling of the type of 
Fig.\,\ref{ssbar_direct} and the subleading 
$\bar K^{(\ast)}  K^{(\ast)}$ 
loops of the type of Fig.\,\ref{ssbar_loop}.

Our scheme has the advantage that it allows
for possible generalizations to non-ideal mixing angles and
different values of the meson decay constants in the strange and non-strange
sectors, respectively.

\section*{Acknowledgements}
One of us (M.N.) appreciates illuminating correspondence with Professor 
G. 't Hooft about the extraction of the mixing angle and the sensitivity to the
various input-schemes of the model used in Section~\ref{eightfold}.  
Work partly supported
by CONCyTEG, Mexico under contract 00-16-CONCyTEG-CONACyT-075.

\newpage

\newpage
\begin{table}
\caption[]{\label{input}
Input (underlined and marked by $^\ast$), 
parameters,  post- and predictions of model 1, 2,
and 3~(a~or~b) which follow the input strategy
of Refs.\,\cite{napsu}, \cite{tornq}, and \cite{thooft3}, respectively.}
{\small
\begin{tabular}{crrrr}
            & model 1  & model 2 & model 3a & model 3b  \\ \hline
$m_\pi$ [Mev]     &  \underline{138 $\pm$ 2 $^\ast$}    & 
 \underline{138 $\pm$ 2 $^\ast$} 
            &  \underline{138 $\pm$ 2 $^\ast$}       & 
\underline{138 $\pm$ 2 $^\ast$} \\ 
$m_K$  [MeV]     &  \underline{496 $\pm$ 3 $^\ast$}    & 
\underline{496 $\pm$ 3 $^\ast$}  
            &  515 $\pm$ 8           & 550 $\pm$ 12 \\
$f_{\pi^0}$ [MeV]     &   \underline{92$\pm$ 4 $^\ast$} &  
\underline{92$\pm$ 4 $^\ast$}
            &   \underline{92$\pm$ 4 $^\ast$} &  
\underline{92$\pm$ 4 $^\ast$}\\
              
$f_{K^0}$ [MeV]  
            &  127 $\pm$ 5       &   \underline{113 $\pm$ 5 $^\ast$}
            &  108 $\pm$ 8        & 101 $\pm$ 6  \\
$m_\eta$ [MeV]   &  \underline{547 $\pm$ 3 $^\ast$}      &  539 $\pm$ 12
            &  \underline{547 $\pm$ 3 $^\ast$}      &  
\underline{547 $\pm$ 3$^\ast$} \\
$m_{\eta'}$ [MeV]&  \underline{958 $\pm$ 3 $^\ast$}     & 963 $\pm$ 11
 &  \underline{958  $\pm$ 3 $^\ast$}  &  \underline{958 $\pm$ 3 $^\ast$} \\
$\frac{1}{2}\sqrt{m_\eta^2\mbox{+}m_{\eta'}^2}$ [MeV]
            & 552 $\pm$ 3             & \underline{552 $\pm$ 3 $^\ast$}
            & 552 $\pm$ 3            & 552 $\pm$ 3\\
$m_{a_0}$ [MeV]  & 919  $\pm$ 4            & 1029 $\pm$ 198
            &   1163  $\pm$ 99                  &   1703   $\pm$ 211\\
$m_{\kappa}$ [MeV]& 927 $\pm$ 7             & 1125 $\pm$ 255
            &  1305    $\pm$ 113               & 1879     $\pm$ 207   \\
$m_{\eta_{\rm ns}}$ [MeV]
            & 851   $\pm$ 7           & 813 $\pm$ 41
            & 778   $\pm$ 17           & 707 $\pm$ 17 \\
$m_{\eta_{\rm s}}$ [MeV]
            & 702  $\pm$ 4            & 746 $\pm$ 41
            & 782  $\pm$ 17            & 847 $\pm$ 15\\
$m_{\eta_{\rm s-ns}}^2$ [MeV$^2$]
            & (535 $\pm$ 5)$^2$     & (560 $\pm$ 21)$^2$
            & (556 $\pm$ 5)$^2$     & (538 $\pm$ 11)$^2$\\
$f_{\eta_{\rm s}}$  [MeV]
            & 161 $\pm$ 7   &   134 $\pm$ 14
                   & 124 $\pm$ 12   &   109 $\pm$ 8\\ 
$a$ [MeV]   & 65 $\pm$ 3         & 65 $\pm$ 3 
            & 65 $\pm$ 3         & 65 $\pm$ 3 \\
$x=\frac{1}{2}\left(\frac{b}{a}-1\right)$ 
        & 0.37 $\pm$ 0.03                     & 0.22 $\pm$ 0.1  
        & 0.17 $\pm$ 0.03                     & 0.09 $\pm$ 0.02\\
$\lambda$   & 14 $\pm$ 2                  & 47  $\pm$ 52 
            & 87 $\pm$ 38                 & 281 $\pm$ 114\\
$\beta$ [MeV]    & $-$1551 $\pm$ 72          & $-$1698 $\pm$ 104 
            &   $-$1672 $\pm$ 99            & $-$1566 $\pm$ 132\\ 
$\xi$  [MeV$^2$]     & (558 $\pm$ 10)$^2$       & (375 $\pm$ 170)$^2$
            &$-$(246 $\pm$ 36)$^2$       & $-$(969 $\pm$ 133)$^2$ \\ 
$\phi_P$ & 56.0 $\pm$ 0.5 $^\circ$         & 49.7 $\pm$ 5.7 $^\circ$
            & 44.7 $\pm$ 2 $^\circ$         & 34.7 $\pm$ 2$^\circ$\\
$\theta_P$ 
            & 1.3 $\pm$ 0.5$^\circ$          &$-$5.0 $\pm$ 5.7 $^\circ$
            & \underline{$-$10 $\pm$ 2 ${^\circ\,}^\ast$} &  
\underline{$-$20 $\pm$ 2 ${^\circ\,}^\ast$}\\
$g_{\eta NN}$ & 2.8 $\pm$ 0.5       & 3.3 $\pm$ 0.6 
              & 3.5 $\pm$ 0.7       & 3.9 $\pm$ 0.7 \\
$g_{\eta'NN}$ & 2.8 $\pm$ 0.4       & 2.3 $\pm$ 0.6 
              & 2.0 $\pm$ 0.7         & 1.2 $\pm$ 0.7 \\
$(g_{\eta NN}^2\mbox{+}g_{\eta'NN}^2)/4\pi$
              & 1.2 $\pm$ 0.4       & 1.3 $\pm$ 0.4 
              & 1.3 $\pm$ 0.4       & 1.3 $\pm$ 0.4 \\
\end{tabular}
}
\end{table}

\newpage
\begin{thebibliography}{00}

\bibitem{thooft1} G.~'t Hooft,
Phys.\ Rev.\ Lett.\  {\bf 37} (1976) 8;
Phys.\ Rev.\  {\bf D14} (1976) 3432.

\bibitem{Okubo} S.~Okubo,
Phys.\ Lett.\  {\bf 5} (1963) 165.

\bibitem{Zweig}
G.~Zweig,
CERN Report No 8419 TH 412, 1964.

\bibitem{Iizuka}
J.~ Iizuka, 
Prog.\ Theor.\ Phys.\ Suppl.\ {\bf 37-38} (1966) 21.
 

\bibitem{MKPRD} M.~Kirchbach,
Phys.\ Rev.\  {\bf D58} (1998) 117901.

\bibitem{Holger} C.~D.~Froggatt and H.~B.~Nielsen,
Origin of symmetries,  World Scientific, (Singapore, Singapore, 1991). 


\bibitem{DonGolHol92}
J.~F.~Donoghue, E.~Golowich and B.~R.~Holstein,
Dynamics of the standard model,
Cambridge monographs on particle physics, nuclear physics and cosmology, 2,
(Cambridge Univ. Pr., Cambridge, 1992).

\bibitem{PDG} The Review of Particle Properties, D.~E.~Groom {\it et al}., 
                 Eur.\  Phys.\ J.\  {\bf C15} (2000) 1.


\bibitem{Feldmann1} T.~Feldmann, P.~Kroll and B.~Stech,
Phys.\ Rev.\  {\bf D58} (1998) 114006,
hep-ph/9802409.

\bibitem{Feldmann2}
T.~Feldmann, 
Int.\ J.\ Mod.\ Phys.\  {\bf A15} (2000) 159,
hep-ph/9907491.

\bibitem{thooft2} G.~'t Hooft,
Phys.\ Rept.\  {\bf 142} (1986) 357.


\bibitem{veneziano}
G.~Veneziano,
Nucl.\ Phys.\ B {\bf 159}, 213 (1979).

\bibitem{witten}
E.~Witten,
Nucl.\ Phys.\ B {\bf 156}, 269 (1979).
\bibitem{schechter}
C.~Rosenzweig, J.~Schechter and C.~G.~Trahern,
Phys.\ Rev.\ D {\bf 21}, 3388 (1980).

\bibitem{arnowitt}
P.~Nath and R.~Arnowitt,
Phys.\ Rev.\ D {\bf 23}, 473 (1981).

\bibitem{diakonov}
D.~Diakonov and M.~I.~Eides,
Problem,''
Sov.\ Phys.\ JETP {\bf 54}, 232 (1981)
[Zh.\ Eksp.\ Teor.\ Fiz.\  {\bf 81}, 434 (1981)].

\bibitem{diakonov2}
D.~Diakonov,
hep-ph/9802298.


\bibitem{luna}
J.~L.~Lucio and M.~Napsuciale,
in:  S. Bianco et al. (Eds.),
Proc. III Workshop on Physics
and Detectors for Daphne (Daphne 99), 
Frascati Physics Series Vol. XVI, 2000, p.~591,
hep-ph/0001136.

\bibitem{nalu} 
J.~L.~Lucio M.\  and M.~Napsuciale,
Phys.\ Lett.\  {\bf B454} (1999) 365,
hep-ph/9903234.

\bibitem{lenaghan}
J.~T.~Lenaghan, D.~H.~Rischke and J.~Schaffner-Bielich,
Phys.\ Rev.\ D {\bf 62}, 085008 (2000), nucl-th/0004006.


\bibitem{thooft3} G.~'t Hooft,
hep-th/9903189.

\bibitem{napsu} M.~Napsuciale,
hep-ph/9803396.

\bibitem{tornq}  N.~A.~Tornqvist,
Eur.\ Phys.\ J.\  {\bf C11} (1999) 359,
hep-ph/9905282.


\bibitem{schshur} T.~Sch\"afer and E.~Shuryak,
hep-lat/0005025.  


\bibitem{isgur}  N.~Isgur and H.~B.~Thacker,
hep-lat/0005006.


\bibitem{Lip} H.~J.~Lipkin,
Nucl.\ Phys.\  {\bf B244} (1984) 147;
ibid.\
{\bf B291} (1987) 720.
      


\bibitem{LipZou} H.~J.~Lipkin and B.~Zou,
Phys.\ Rev.\  {\bf D53} (1996) 6693.

\bibitem{Zou}
B.~Zou,
Phys.\ Atom.\ Nucl.\  {\bf 59} (1996) 1427, 
hep-ph/9611238.



\bibitem{proton}
D.~Adams {\it et al.}  [Spin Muon Collaboration (SMC)],
Phys.\ Rev.\  {\bf D56} (1997) 5330, hep-ex/9702005.

\bibitem{ShoVe} G.~M.~Shore and G.~Veneziano, Phys.\ Lett.\ {B244}, 
(1990) 75; Nucl.\ Phys.\ {B381} (1992) 23.

\bibitem{GrKr}  W.~Grein and P.~Kroll,
Nucl.\ Phys.\  {\bf A338} (1980) 332.

\bibitem{MAMI} 
B.~Krusche {\it et al.},
Phys.\ Rev.\ Lett.\  {\bf 74} (1995) 3736.

\bibitem{TiKa} 
    L.~Tiator, C.~Bennhold and S.~S.~Kamalov,
    Nucl.\ Phys.\  {\bf A580} (1994) 455, nucl-th/9404013.

\bibitem{KDO} H.~Kim, T.~Doi, and M.~Oka,
Nucl.\ Phys.\ {\bf A662}, 371 (2000),
nucl-th/9909007.

\bibitem{Zhu} S.-L.~Zhu, Phys.\ Rev.\ {\bf C61}, 065205 (2000),
nucl-th/0002018.

\bibitem{Saghai} Q.~Zhao, B.~Saghai, and Zh.~Li, nucl-th/0011069.

\bibitem{Moskal}
P.~Moskal {\it et al.},
Phys.\ Rev.\ Lett.\  {\bf 80} (1998) 3202,
nucl-ex/9803002;
AIP Conf.\ Proc.\  {\bf 512} (2000) 65, nucl-ex/0007002.

\bibitem{DeGrand}
T.~DeGrand and A.~Hasenfratz,
hep-lat/0103002,
hep-lat/0012021,

\bibitem{isgur2}
I.~Horvath, N.~Isgur, J.~McCune and H.~B.~Thacker,
hep-lat/0102003.



\bibitem{KaiLeut} 
R.~Kaiser and H.~Leutwyler,
Eur.\ Phys.\ J.\ C {\bf 17} (2000) 623,
hep-ph/0007101.




\end{thebibliography}
\end{document}